\documentclass[]{article}
\usepackage{graphicx}
\usepackage{subfig}
\usepackage{color}
\usepackage[colorlinks=true,linkcolor=blue]{hyperref}
\usepackage{url}
\usepackage[normalem]{ulem}
\usepackage{cancel}
\usepackage{cite}
\usepackage[dvipsnames]{xcolor}
\usepackage{booktabs}
\usepackage[pagewise]{lineno}
\usepackage{hyperref}
\usepackage{comment}
\usepackage{verbatim}
\usepackage{caption}
\usepackage[normalem]{ulem}
\usepackage{amsmath,amssymb,mathrsfs}
\usepackage[margin=0.8in]{geometry}

\usepackage[affil-it]{authblk} 
\usepackage{etoolbox}
\usepackage{lmodern}

\title{Towards enhanced performance in fusion plasmas via turbulence suppression by MeV ions}

\author{S. Mazzi$^{1,2,*}$, J. Garcia$^{2,**}$, D. Zarzoso$^{1}$, Ye.O. Kazakov$^{3}$, J. Ongena$^{3}$, 
M. Nocente$^{4,5}$, M. Dreval$^{6}$, \v{Z}. \v{S}tancar$^{7}$, G. Szepesi$^{8}$, J. Eriksson$^{9}$, A. Sahlberg$^{9}$, 
S. Benkadda$^{1}$ and JET contributors$^{a}$}

\affil{$^1$Aix-Marseille Université, CNRS PIIM, UMR 7345 Marseille, France\\
	$^2$CEA, IRFM, F-13108 Saint-Paul-lez-Durance, France\\
	$^3$Laboratory for Plasma Physics, LPP-ERM/KMS, EUROfusion Consortium member, TEC Partner, Brussels, Belgium\\
	$^4$Dipartimento di Fisica “G. Occhialini”, Università di Milano-Bicocca, Milan, Italy\\
	$^5$Institute for Plasma Science and Technology, National Research Council, Milan, Italy\\
	$^6$National Science Center Kharkiv Institute of Physics and Technology, 1 Akademichna Str., Kharkiv 61108, Ukraine\\
	$^7$Jožef Stefan Institute, Jamova cesta 39, SI-1000 Ljubljana, Slovenia\\
	$^8$CCFE, Culham Science Centre, Abingdon, Oxon, OX14 3DB, United Kingdom\\
	$^{9}$Department of Physics and Astronomy, Uppsala University, Uppsala, Sweden\\
	$^a$See the author list of E. Joffrin et al., Nucl. Fusion 59, 112021 (2019)\\
	\vspace{1em}
	$^*$\textbf{Email}: samuele.mazzi@univ-amu.fr\\
	$^{**}$\textbf{Email}: jeronimo.garcia@cea.fr}

\date{}

\begin{document}
	
\maketitle
	
\begin{abstract}
Megaelectron volt (MeV) alpha particles will be the main source of plasma heating in magnetic confinement fusion reactors. Yet, instead of heating fuel ions, most of the energy of alpha particles is transferred to electrons. Furthermore, alpha particles can also excite Alfvénic instabilities, previously considered as detrimental. Contrary to expectations, we demonstrate efficient ion heating in the presence of MeV ions and strong fast-ion driven Alfvénic instabilities in recent experiments on the Joint European Torus (JET). Detailed transport analysis of these experiments with state-of-the-art modeling tools explains the observations. Here we show a novel type of turbulence suppression and improved energy insulation in plasmas with MeV ions and fully developed Alfvénic activities through a complex multi-scale mechanism that generates large-scale zonal flows. This mechanism holds promise for a more economical operation of fusion reactors with dominant alpha particle heating and, ultimately, cheaper fusion electricity.
		
\end{abstract}
\vspace{2em}

The urgent need for environmentally friendly energy sources becomes increasingly more important for the future of our modern society. Among several options, the fusion reaction between the hydrogen isotopes deuterium (D) and tritium (T) holds the promise for a safe clean and inexhaustible energy production: $\mathrm{D} + \mathrm{T}$ $\rightarrow$ $^4\mathrm{He}$ (3.5 $\mathrm{MeV}) + n$ (14.1 MeV), with an alpha particle ($^4\mathrm{He}$) and a neutron ($n$) as fusion products. This reaction needs temperatures around 150 million K and can be realized in magnetic-confinement fusion devices \cite{ongena2016magnetic}. The largest fusion device currently in operation is the Joint European Torus (JET) \cite{litaudon2017overview}. An even larger experimental device ITER has started its assembly phase in July 2020, aiming to demonstrate the scientific and technological matureness of magnetic confinement fusion \cite{shimada2007progress}.

\begin{figure*}[htb!] 
    \centering
	\subfloat{\includegraphics[scale=0.44]{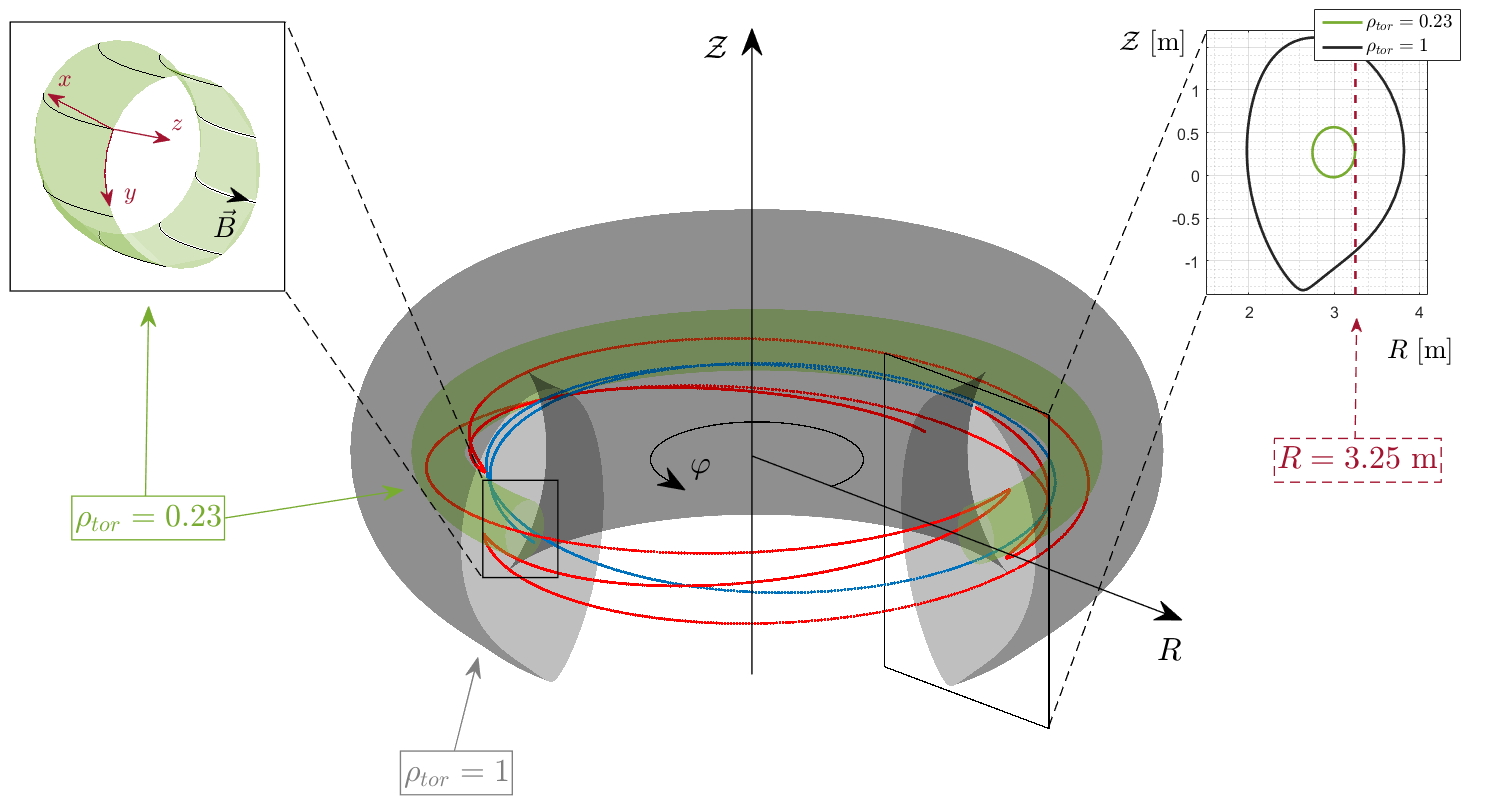}}
	\caption{\textbf{Schematic view of the tokamak geometry}. Charged particles, confined by a combination of toroidal and poloidal magnetic fields, can either circulate along the magnetic field lines in the toroidal direction $\varphi$ (so-called passing particles, blue trajectory), or periodically reverse their parallel velocity (so-called trapped particles, red trajectory). In this drawing of the particle orbits, the rapid gyro-motion around the field line is averaged out, and only the guiding centre motion is shown. The magnetic flux surface corresponding to the radial coordinate $\rho_{tor}=0.23$ is displayed in green. The inset at the top on the left illustrates the field-aligned set of coordinates used in the flux-tube version of the \textsc{Gene} code, with $(x,y,z)$ the radial, binormal and parallel coordinate, respectively. The figure at the top on the right shows a poloidal cross-section of the plasma in the ($R$,$\mathcal{Z}$) plane, where $R$ is the radial distance from the torus centre and $\mathcal{Z}$ is the vertical coordinate; in green the flux surface for $\rho_{tor}=0.23$ and in black that for $\rho_{tor}=1$, corresponding to the plasma boundary, are shown.}
	\label{Tokamak}
\end{figure*}

Both JET and ITER are based on the tokamak concept, a toroidal configuration schematically shown in \linebreak Figure \ref{Tokamak}. 
Large temperature gradients ($\sim$100 million K per meter) are unavoidably present in these devices because of the required high plasma temperature in the center and the necessity for a cold plasma edge to avoid damage of the wall of the device. These gradients create instabilities and turbulence, often with very different spatio-temporal scale lengths \cite{doyle2007plasma}, that result in energy and particle transport. One of the most important is the ion temperature gradient (ITG) instability with a characteristic scale length $\sim$10$^{-3}$ m, orders of magnitude smaller than the plasma size ($\sim$1 m). The microscopic ITG instability largely limits plasma temperatures that can be achieved in a fusion device \cite{romanelli1989ion}.

The success of magnetic confinement fusion as an energy source relies crucially on reaching high temperatures for the D and T ions. Fusion-born alpha particles are the main source of central plasma heating in ITER and future fusion power plants. Yet, these highly energetic alpha particles heat primarily electrons rather than bulk ions through Coulomb collisions. The physics of plasma heating by alpha particles is complex and its extrapolation to future devices is not straightforward, partly because of the mutual interplay between turbulence and energetic ions. This can lead to additional nonlinearities \cite{zarzoso2013impact}, further increasing the complexity of plasma dynamics. As an example, a reduction of the ITG-induced transport was recently predicted for typical expected ITER D-T plasmas in the presence of fusion-born alpha particles \cite{garcia2018isotope}.
	
Simultaneously with plasma heating, alpha particles can also excite instabilities, in particular, the non-damped Alfvén eigenmodes (AEs). The excitation of these modes by energetic ions can be related to the particular geometry of toroidal fusion plasmas such as the toroidicity (toroidicity-induced AEs or TAEs)\cite{cheng1985high,cheng1986low} and the non-circular plasma cross-section (ellipticity-induced AEs or EAEs)\cite{betti1991ellipticity}. The presence of these modes is commonly considered detrimental for plasma confinement in fusion plasmas and ITER, as they can cause increased transport of energetic ions \cite{fasoli2007physics,gorelenkov2014energetic,todo2019introduction,heidbrink2020mechanisms}.

The nonlinear three-way interaction between MeV-range ions, fast-ion driven AEs and microturbulence has not been systematically studied at ITER-relevant conditions, either theoretically or experimentally. The last full-scale experimental D-T campaign dates back to 1997 \cite{keilhacker1999high}, also highlighting the long-standing puzzle known as \textit{anomalous} ion heating by alpha particles \cite{thomas1998observation}. Preliminary studies \cite{testa2012phenomenological} suggested that such an increase of ion temperature in JET D-T plasmas was actually not attributable to direct alpha particle heating, but to ion confinement enhancement. Since then a tremendous progress in understanding fusion plasma turbulence and in the development of first-principle turbulence codes has been reached \cite{fasoli2016computational}. 

Controlling turbulence in a fusion power plant will finally lead to a more economical operation and ultimately to a reduction in the cost of fusion electricity. This paper reports on a novel mechanism of turbulence suppression in the presence of a significant population of MeV range fast ions and fully destabilized fast-ion driven TAEs, identified during detailed transport analysis of JET experiments in D-$^3$He plasmas. The improved confinement in JET plasmas with MeV-range fast ions was confirmed with state-of-the-art turbulence modeling. Thermal ion energy fluxes perpendicular to the magnetic surfaces were suppressed because of the appearance of intense poloidally directed shear flows, known as zonal flows \cite{diamond2005zonal}. Zonal flows, which are solely generated by nonlinear interactions, are a common phenomenon in nature and can be remarkably stable as, e.g., the famous belts of Jupiter \cite{heimpel2005simulation}. In this paper, turbulence modeling shows that these zonal flows, in both electrostatic and electromagnetic field perturbations, are triggered as a result of the nonlinear interaction between MeV-range ions, fully destabilized fast-ion driven TAEs and microturbulence. Thus, these zonal flows regulate and suppress ion-scale microturbulent transport, increasing the thermal insulation of the central plasma.  

Unlike previous numerical analyses of low-energetic fast ions triggering marginally stable AEs \cite{di2019electromagnetic}, the results here reported highlight that unstable AEs - excited by much more energetic fast ions - can be, in fact, beneficial to improve thermal energy confinement, a concept that has not been extensively pursued before.
A detailed understanding of this complex interplay paves the way towards controlling turbulence and enhanced performance of future fusion reactors with strong alpha particle heating.

\subsection*{Improved confinement in JET plasmas with a large population of MeV-range fast ions}
The impact of fast ions on plasma dynamics was studied in JET D-$^3$He plasmas ($n_{^3\mathrm{He}}/n_e \approx$ 20-30\%). Although the presence of TAEs is usually accompanied by a loss of energy and particle confinement \cite{chen2016physics,breizman2011major,fasoli2007physics,gorelenkov2014energetic,heidbrink2008basic}, we show that the energy confinement can be actually improved in the presence of these modes. This is illustrated by comparing two JET pulses performed at the same plasma conditions, differing only in the auxiliary heating scheme.

\begin{figure*}[hbt!]
	\centering
	\subfloat{\includegraphics[scale=0.2]{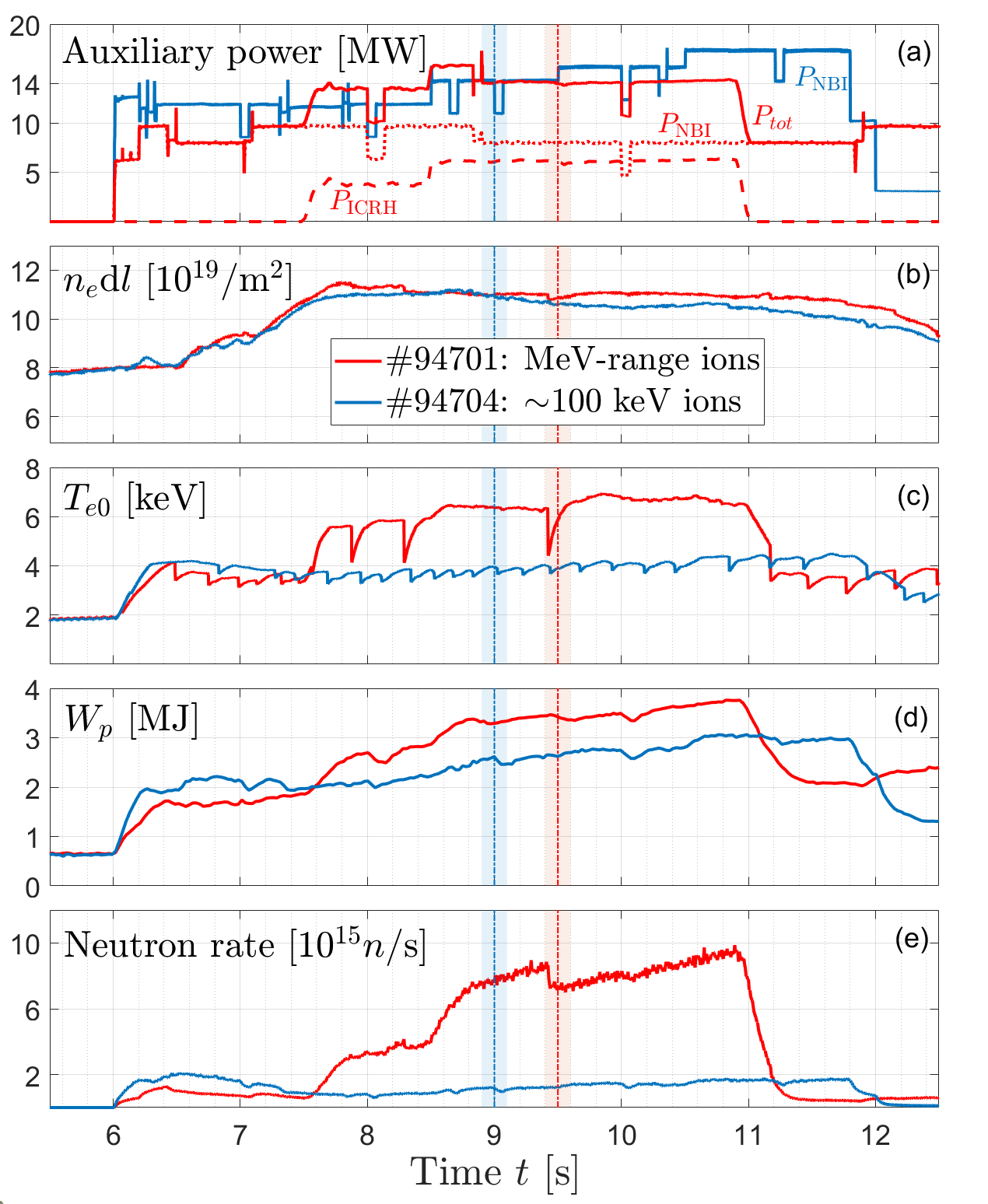}}
	\subfloat{\includegraphics[scale=0.2]{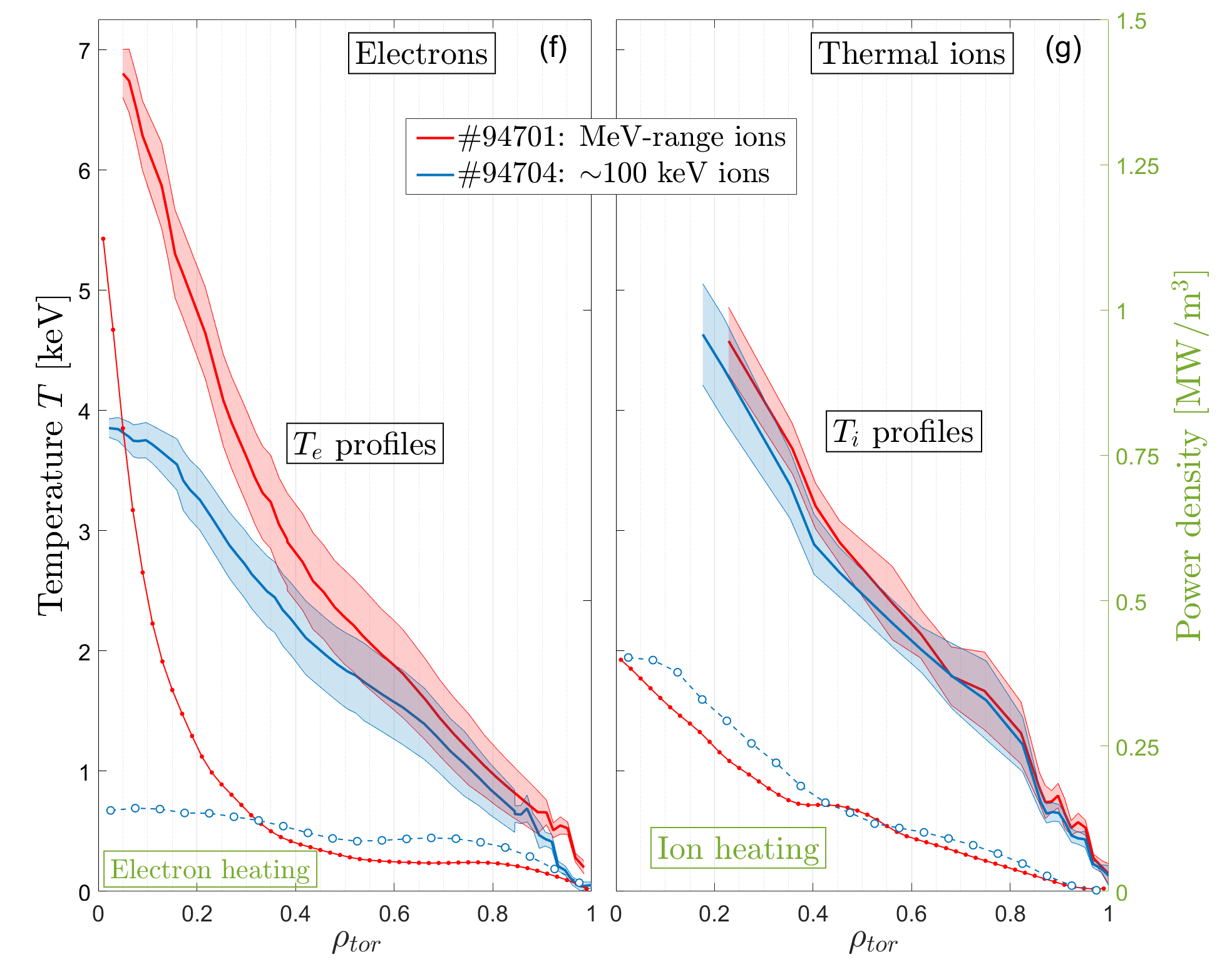}}
	\caption{\textbf{Overview of the main parameters in JET pulses \#94701 and \#94704}. The panels show (a) the auxiliary heating power $P_{aux}$ from the NBI and ICRH systems ($P_{tot}=P_{\mathrm{NBI}}+P_{\mathrm{ICRH}}$), (b) core line integrated electron density ($\int n_e \mathrm{d}l$), (c) core electron temperature $T_{e0}$, (d) plasma stored energy $W_p$ and (e) neutron rate from D-D fusion reactions. Red traces correspond to pulse $\#$94701, blue traces to pulse $\#$94704. The vertical lines show the times used in (f) and (g) for the electron and ion temperature profile comparison. The chosen timings ($t=9.5$ s for \#94701 and $t=9$ s for \#94704) are taken at the same total input power ($14$ MW) and very similar plasma densities for both discharges. In (f) and (g), also the deposition profiles of the total heating power on electrons and thermal ions as computed by TRANSP for both pulses are shown. In the power deposition calculations both the externally heating power (NBI + ICRH) and the electron/ion energy exchanges are considered.}
	\label{Waveforms}
\end{figure*}

In pulse \#94704 (blue lines in Figure \ref{Waveforms}), Neutral Beam Injection (NBI) was the only heating system \linebreak ($P_{aux} =$ 12-16 MW), injecting fast D ions with energies up to 100 keV. The measured temperature profiles of electrons and ions in the phase with $P_{aux} = 14$ MW ($t = 9.0$ s) are shown in Figures \ref{Waveforms}(f) and (g), together with the NBI power deposition profiles to bulk ions and electrons, as computed by the Monte-Carlo orbit following code NUBEAM \cite{pankin2004tokamak} implemented in the modelling suite TRANSP \cite{ongena2012numerical}. The moderately energetic D ions from NBI deposit most of their energy to bulk ions, leading to $T_i/T_e \approx 1.4$ measured at $\rho_{tor}=0.2$ ($\rho_{tor}$ is the normalized toroidal flux coordinate). No fast-ion driven AEs were observed in this plasma.

\begin{figure*}[hbt!]
	\subfloat{\includegraphics[scale=0.2]{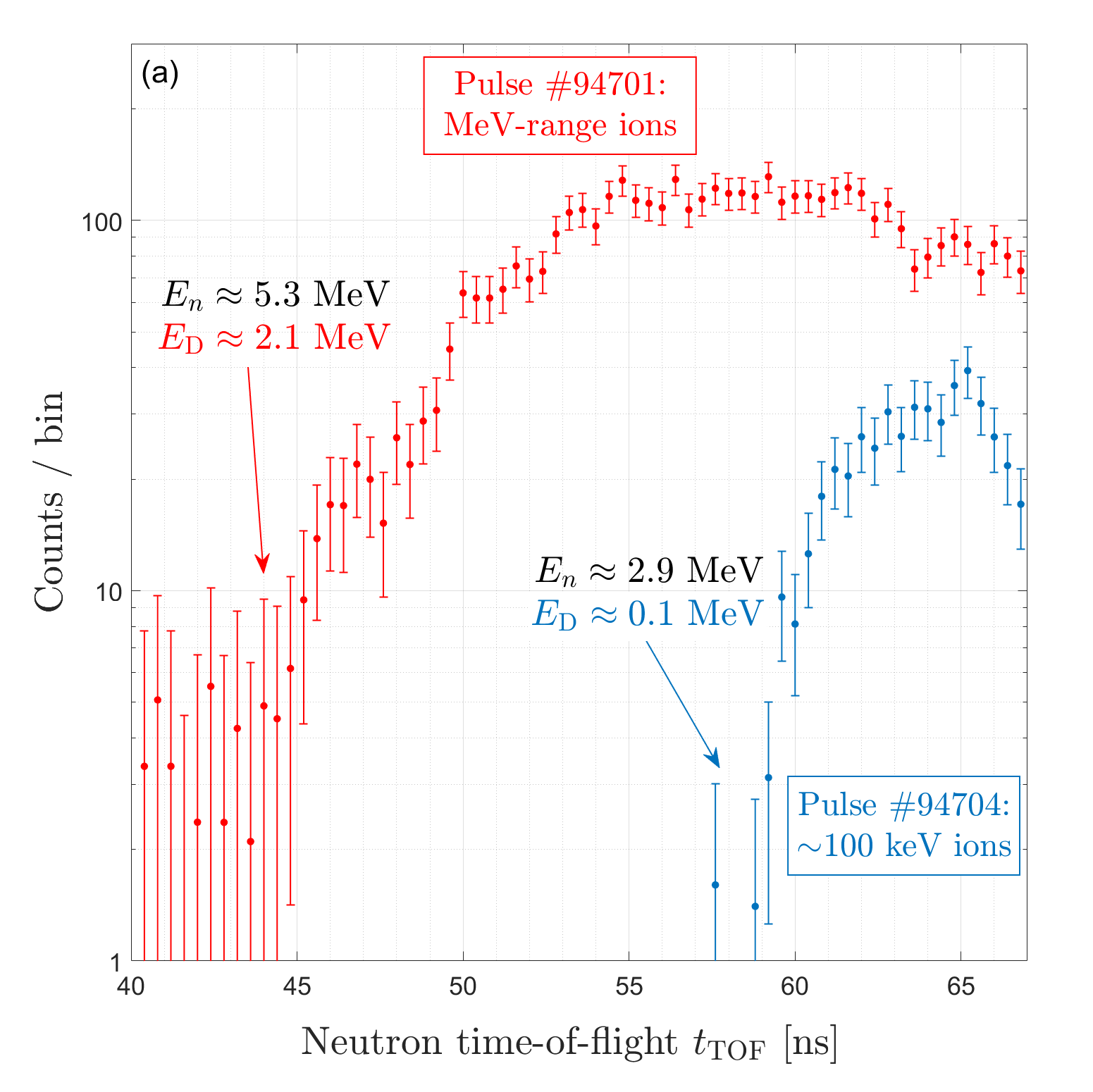}}
	\subfloat{\includegraphics[scale=0.54]{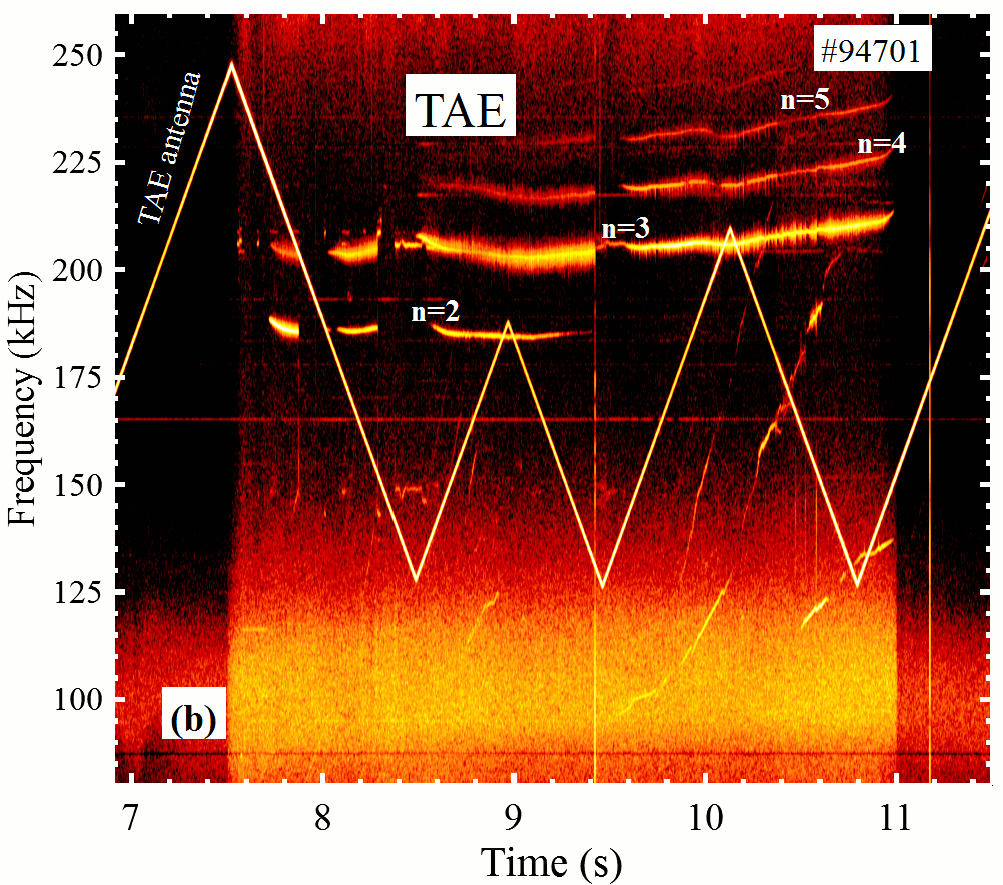}}
	\caption{\textbf{Evidence of MeV-range fast D ions and fast-ion driven TAEs}. Panel (a) shows TOFOR \cite{johnson20082} measurements of the time-of-flight of the neutrons for both \#94701 and \#94704, indicating the presence of MeV-range fast D ions generated by the three-ion scheme. The time-of-flight $t_{\mathrm{TOF}}$ of the D-D fusion-born neutrons can be related to the energy of the reactant fast D ions by kinematics considerations \cite{jacobsen2017velocity}. In (b), the Mirnov coil spectrograms for pulse $\#$94701. Strong TAE activity in the range $185$-$235$ kHz during the application of the three-ion heating scheme is observed. Other Alfvén eigenmodes, such as EAEs and RSAEs, have been detected as well \cite{nocenteNF2020}. Toroidal mode numbers are indicated next to the respective traces of the detected magnetic fluctuations in the spectrogram.}
	\label{MirnovCoils94701}
\end{figure*}

In pulse \#94701 (red lines in Figure \ref{Waveforms}), D ions from NBI were accelerated to higher energies with waves in the ion cyclotron range of frequencies (ICRF). The three-ion scenario using a combination of 8 MW of NBI and 6 MW of ICRF produced D ions with tail energies up to $\sim$2-3 MeV in the plasma core \cite{kazakov2017efficient,ongena2017synergetic,kazakov2020plasma,nocenteNF2020}. This is demonstrated in Figure \ref{MirnovCoils94701}(a), where measurements from the TOFOR spectrometer \cite{johnson20082} are shown. A rich variety of TAEs with different toroidal mode numbers were excited, see Figure \ref{MirnovCoils94701}(b). The radial location of TAE modes was estimated to be in the range $3.2$ m $< R < 3.35$ m, inferred from X-mode correlation reflectometer measurements available in a very similar pulse (see Supplementary information). As a result of their collisional slowing-down, these highly energetic D ions heat predominantly electrons, giving rise to a strongly peaked $T_e$ profile. This is illustrated for $t = 9.5$ s in Figure \ref{Waveforms}(f), also at $P_{aux} =14$ MW. Surprisingly, $T_i$ is also higher in pulse \#94701 although the power deposition to bulk ions in the core is lower because of the reduced NBI power, compared to pulse \#94704.

Electron heating has been always considered a drawback with respect to ion heating by NBI as it also leads to a reduction of well-known mechanisms stabilizing ITG turbulence such as the ratio $T_i/T_e$ and the $E \times B$ shearing rate ($\gamma_{E \times B}$) \cite{doyle2007plasma}. For the pulses considered in this paper, at the radial location $\rho_{tor}=0.23$, where high-quality $T_i$ measurements are available, such parameters are significantly lower for pulse \#94701 ($T_i/T_e \approx 1$ and \linebreak $\gamma_{E \times B} \approx 1.42$ $\mathrm{ms}^{-1}$) compared to pulse \#94704 ($T_i/T_e \approx 1.4$ and $\gamma_{E \times B} \approx 2.15$ $\mathrm{ms}^{-1}$). Therefore, unlike the experimental results obtained, lower confinement could have been expected for the pulse \#94701 compared to \#94704.

However, these two pulses also exhibit a large difference in the fast-ion energy and density. This can readily be seen by comparing the measured neutron rate at the same amount of auxiliary heating power: \linebreak 8-9 $\times 10^{15}$ $\mathrm{s}^{-1}$ in \#94701 vs. $1.2 \times 10^{15}$ $\mathrm{s}^{-1}$ in \#94704, see Figure \ref{Waveforms}(d). This is further evidenced in Figure \ref{MirnovCoils94701}(a) \linebreak that illustrates neutron energy spectra in these two pulses at ${P_{aux} = 14}$ MW, measured by the time-of-flight neutron spectrometer TOFOR \cite{johnson20082}. Note that shorter times-of-flight correspond to higher neutron energies,\linebreak $E_n \approx 2.9$ MeV $\times $(60 $\mathrm{ns}/t_{\mathrm{TOF}})^2$ and this reflects the presence of high-energy D ions in the plasma \linebreak (cf. Figure 3(b) in \cite{eriksson2018measuring}). The observation of neutrons with $t_{\mathrm{TOF}} =$ 40-45 ns is a direct experimental confirmation for a significant amount of D ions with energies up to $\sim$2-3 MeV in pulse \#94701. The large difference in the fast-ion content between pulses \#94701 and \#94704 was reproduced in numerical simulations. For pulse \#94704 with $\sim$100 keV ions, TRANSP shows $T_{fast}/T_e \approx 8.7$ and $n_{fast}/n_e \approx 6\%$, comparable to previous studies at JET that demonstrated the beneficial effect of moderately energetic fast ions to reduce turbulence \cite{citrin2013nonlinear,garcia2015key,di2019electromagnetic}. However, with MeV-range ions, these values become $T_{fast}/T_e \approx 33.6$ and $n_{fast}/n_e \approx 3\%$ and are much closer to those expected for fusion-born alpha particles in ITER D-T plasmas \cite{garcia2018isotope}.

\subsection*{Numerical modeling of turbulence in the presence of MeV-range ions}
Extensive numerical analyses of turbulence in JET pulse \#94701 with a large population of MeV-range ions were carried out with the state-of-the-art gyrokinetic code \textsc{Gene} \cite{jenko2000electron}. Applied in its local (also called 'flux-tube') version, the \textsc{Gene} code has been successfully validated and has reproduced turbulence properties for various tokamak plasma experiments (see e.g.~Refs.~\cite{citrin2013nonlinear,citrin2014electromagnetic,garcia2015key,gorler2016validation,mazzi2020impact}).
	
The simulation domain consists of a tube located around the flux surface $\rho_{tor} = 0.23$, corresponding to \linebreak $R=3.25$ m in the mid-plane at the low field side. This region of the plasma is characterized by large temperature gradients together with a significant population of fast ions and the localization of TAE modes. Experimental data at $t = 9.5$ s  were taken as input for \textsc{Gene} simulations (the input parameters are summarized in Table \ref{Table_input_parameters} in the Methods section). An equivalent Maxwellian distribution was used to represent the fast D population in \textsc{Gene}. Note that although it does not fully represent the detailed distribution function, earlier studies showed that it is still a rather good approximation to reproduce the main transport characteristics, including the experimental power balances \cite{di2018non,bonanomi2018turbulent}.
	
Linear stability analysis shows that, as expected, ITG modes dominate the spectrum, peaking at $k_y\rho_s \approx 0.45$. Here, $k_y$ is the binormal (‘poloidal’) wavenumber and $\rho_s$ is the characteristic Larmor radius of thermal ions at the sound speed, $c_s = (T_e/m_p)^{1/2}$, with $m_p$ the proton mass. Because of the uncertainties in the fast-ion profiles calculated by TRANSP in the presence of destabilized AEs, we performed a scan over the normalized fast-ion logarithmic pressure gradient, $R/L_{p_{\mathrm{FD}}}$. The outcome of this assessment is that fast D ions destabilize a mode in the low-$k_y$ region ($k_y\rho_s<0.1$) for sufficiently large values of $R/L_{p_{\mathrm{FD}}}$. This mode was identified as a fast-ion driven TAE since the computed mode frequency ($f \approx 200$ kHz) is very close to the TAE gap of the Alfvén continuum \cite{cheng1986low,fu1989excitation} and the experimentally measured TAE frequencies (Figure \ref{MirnovCoils94701}(b)).

\subsection*{Suppression of the electrostatic turbulence in the presence of fast-ion-driven TAEs}
To study the impact of MeV-range ions and fast-ion driven modes on the turbulence dynamics of the plasma, nonlinear simulations with \textsc{Gene} were performed.
\begin{figure*}[hbt!] 
	\centering
	\subfloat{\includegraphics[scale=0.4]{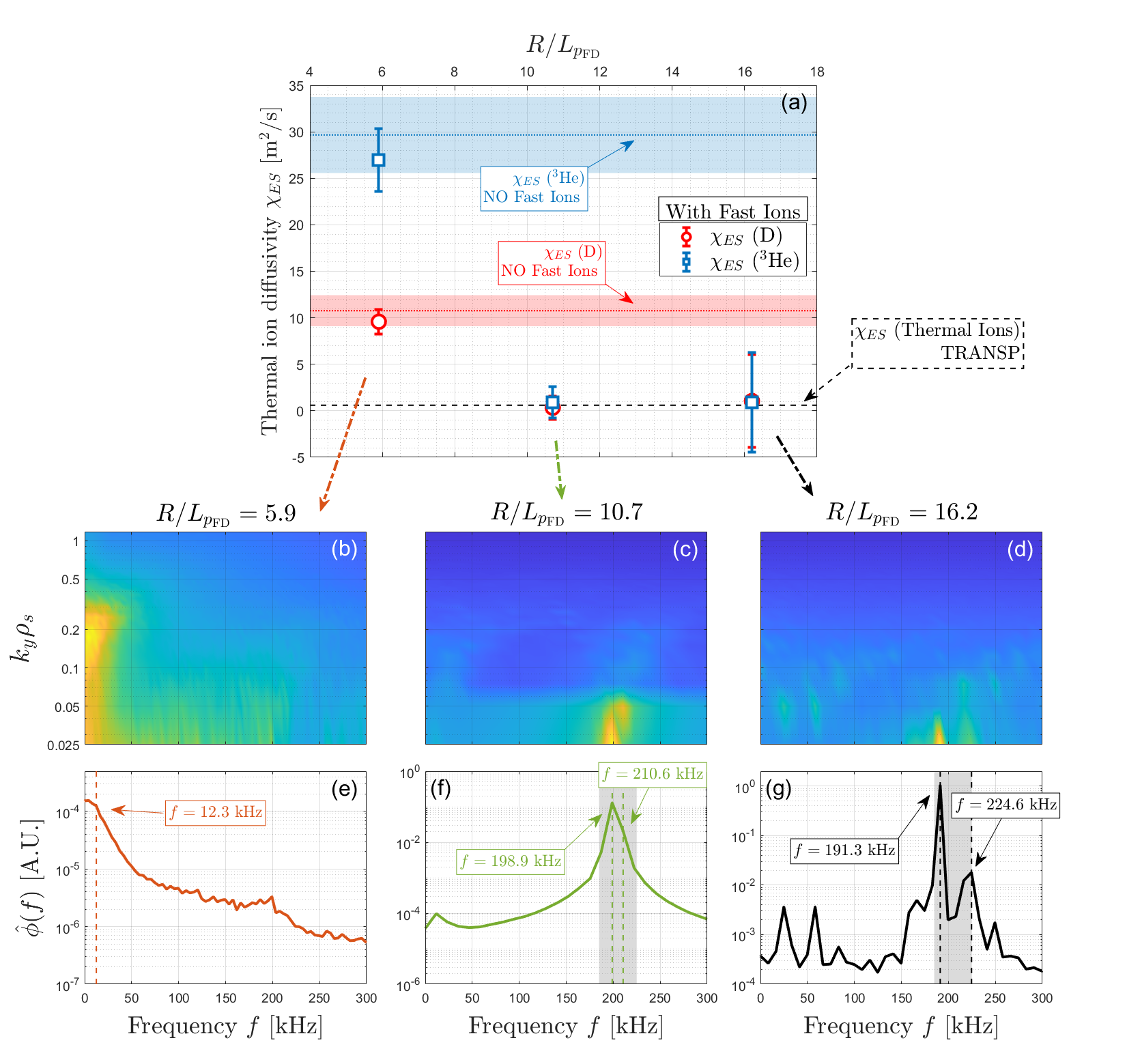}}
	\caption{\textbf{Thermal ion transport suppression simultaneous with the destabilization of FI-driven modes}. In panel (a), the electrostatic thermal diffusivities of D and $^3$He as a function of the fast ion normalized logarithmic pressure gradient $R/L_{p_{\mathrm{FD}}}$ are shown. The error bars represent the standard deviation obtained from the code calculation. The large error bars for the rightmost points of the scan can be attributed to the high-frequency and large-amplitude modulations of the energy fluxes due to strong FI-driven fluctuations of the electrostatic potential. The D and $^3$He thermal diffusivities for the case without fast ions are plotted in horizontal dotted lines within shaded areas representing the standard deviations. The horizontal black dashed line represents the thermal ion energy diffusivity ($\chi_{ES} = 0.6$ $\mathrm{m^2/s}$) obtained with TRANSP. Note the good agreement between the two curves only when the FI-driven TAEs are nonlinearly destabilized. Panels (b-d) show the frequencies of the Fourier transform of the gyroaveraged perturbed electrostatic potential fluctuation ($\hat{\phi}$) as a function of the binormal wavenumber and the frequency and averaged over the other spatial directions ($x$ and $z$). The three cases refer to the points of the scan displayed in panel (a). In (e-g), $\hat{\phi}$ is further averaged over the binormal direction.
	The grey shaded areas in panels (f) and (g) represent the experimentally measured range of TAE frequencies showing good agreement with the frequency peaks.}
	\label{HeD_DiffES}
\end{figure*}
Figure \ref{HeD_DiffES}(a) shows the energy diffusivity attributed to electrostatic (ES) fluctuations for both thermal ion species, $\chi_{ES}(\mathrm{D})$ and $\chi_{ES}(^3\mathrm{He})$. Yet, in case of stronger fast-ion pressure gradients, e.g. $R/L_{p_{\mathrm{FD}}}=10.7$ and $R/L_{p_{\mathrm{FD}}}=16.2$, the nonlinear simulations reveal a strong stabilizing effect of the fast ions on the diffusivity of the thermal ions. Such a stabilizing effect occurs only when the threshold in the fast-ion pressure gradient is exceeded such that fast-ion driven TAEs are excited in the nonlinear turbulence regime (see also Figures \ref{HeD_DiffES}(f) and (g)). In the presence of these TAE modes, the computed thermal diffusivities of both thermal ion species were reduced by more than 95\% compared to the values obtained without fast ions. Note that the strongly reduced values of $\chi_{ES}$ predicted by \textsc{Gene} are in a very good agreement with the estimated diffusivity of thermal ions, as calculated by the power balance analysis in TRANSP based on the experimentally measured temperature profiles (the horizontal black dashed line in Figure \ref{HeD_DiffES}(a)).

To unveil additional information on the impact of fast ions on the turbulence spectrum, Figures \ref{HeD_DiffES}(b)-(d) show the amplitudes of the Fourier components of the perturbed electrostatic potential $\hat{\phi}$ (averaged over the radial and parallel directions) as a function of $k_y\rho_s$ and frequency for three values of $R/L_{p_{\mathrm{FD}}}$. The frequency dependence of $\hat{\phi}$ additionally averaged over $k_y\rho_s$ is illustrated in Figures \ref{HeD_DiffES}(e)-(g). The spectra reveal that for $R/L_{p_{FD}} = 5.9$ the dominant modes are characterized by low frequencies and peak around $k_y\rho_s \approx 0.2$, characteristic for the ITG instability. In contrast, at larger fast-ion gradients a modification in the dominant turbulence structure is observed and the dominant modes are characterized by larger scales (lower $k_y\rho_s$) and shifted towards higher frequencies, $f\approx200$ kHz. The computed mode frequencies are very close to those found in linear \textsc{Gene} simulations and even more importantly to the frequencies of experimentally observed TAEs. 

\begin{figure*}[hbt!]
	\centering
	\subfloat{\includegraphics[scale=0.34]{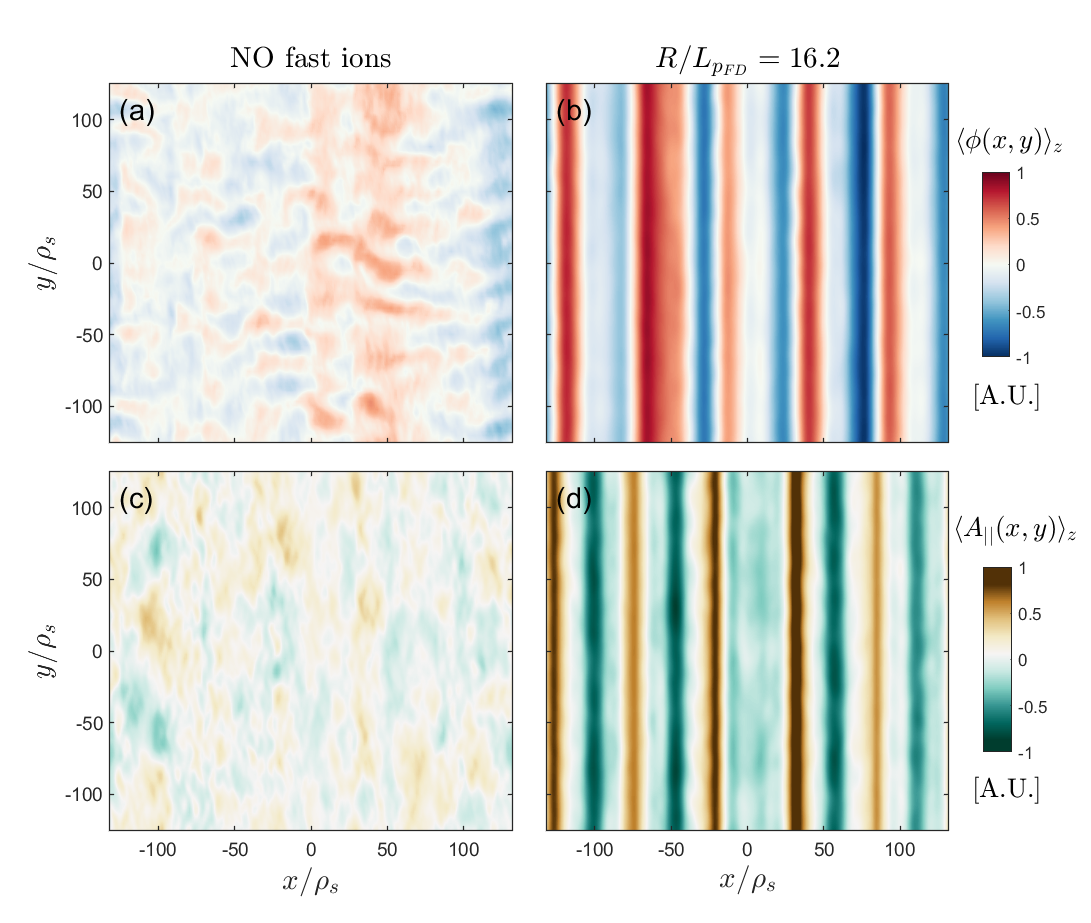}}
	\subfloat{\includegraphics[scale=0.375]{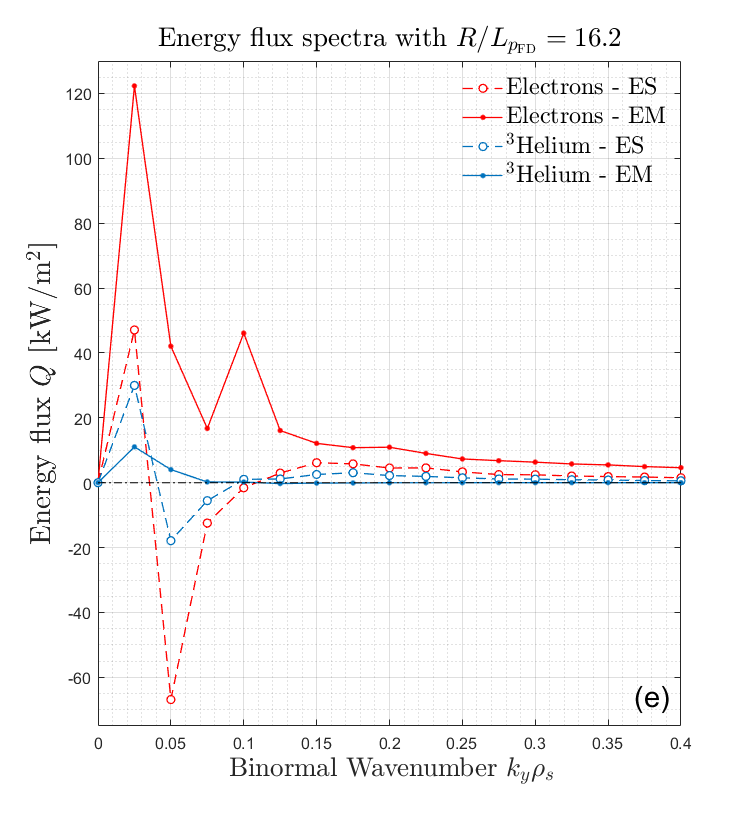}}
	\caption{\textbf{Evidence for the nonlinear generation of zonal flows}. Contour plots of the gyroaveraged perturbed electrostatic potential $\phi$ and the component in the direction of the magnetic field of the potential vector $A_{||}$ in the $(x,y)$ space for (a) and (c) the case without fast ions and (b) and (d) the $R/L_{p_{FD}} = 16.2$ case. A time average is performed for both field fluctuations at the the low field side mid-plane. The $\phi$ contour plots (a) and (b) are both normalized to the maximum (absolute) value of the $R/L_{p_{FD}} = 16.2$ case. The same procedure is applied for $A_{||}$ in plots (c) and (d). This procedure highlights the increase of the intensity of both field perturbations when fast-ion driven TAEs are unstable (for $R/L_{p_{FD}} = 16.2$) with respect to the simulation without fast ions. In (e), the electrostatic (ES) and electromagnetic (EM) energy flux spectra are shown for both electrons and $^3$He with $R/L_{p_{FD}} = 16.2$. Although the $k_y$ simulation domain extends up to $k_y\rho_s=1.2$, the plot is resized for displaying properly the interesting region of the spectra.}
	\label{ContPlotxy_Spectra}
\end{figure*}

With the knowledge of the Fourier components, the perturbed electrostatic potential $\phi$ can be calculated in real space. The surprising results, shown in Figures \ref{ContPlotxy_Spectra}(a) and (b), illustrate the dependence of $\phi$ on the radial ($x/\rho_s$) and binormal ($y/\rho_s$) coordinates.
In the absence of fast ions, one can see small-scale turbulent eddies predominately elongated along the radial direction, typical for ITG-induced transport. This is in strong contrast with the case $R/L_{p_{FD}} = 16.2$, where intense poloidally oriented and radially sheared zonal flows are clearly seen. These zonal flows are well-known to strongly affect the ITG nonlinear saturation \cite{diamond2005zonal} and produce a radial de-correlation of the turbulent eddies \cite{biglari1990influence}, thereby leading to a reduction in the turbulent transport. The importance of zonal flows to suppress ion-scale turbulence was also recently pointed out in view of future ITER operations \cite{candy2016crucial}.
	
The complex interplay between fully destabilized TAEs and the nonlinearly generated zonal flows is responsible for the suppression of the electrostatic component of the thermal ion energy fluxes. The zonal flow shearing rate nearly doubles for $R/L_{p_{FD}}=16.2$, corresponding to the conditions with unstable TAE modes excited by fast ions. Note that the suppression of the electrostatic turbulence in the presence of marginally stable TAEs, appearing in \textsc{Gene} simulations, was discussed in \cite{di2019electromagnetic}. Our results highlight that electrostatic turbulence suppression is also possible in plasmas with fully destabilized TAEs, supported by experimental observations.
	
\subsection*{Mitigation of electromagnetic transport due to zonal field activity}
Because of the strong suppression of the electrostatic thermal fluxes, the plasma turbulent regime becomes purely electromagnetic. Figure \ref{ContPlotxy_Spectra}(e) shows the computed electrostatic and electromagnetic energy flux spectra for electrons and thermal $^3\mathrm{He}$ ions as a function of $k_y\rho_s$ for $R/L_{p_{FD}} = 16.2$. Among these, the electromagnetic electron transport strongly dominates, mainly induced by the low-$k_y$ fluctuations of the magnetic field ($k_y\rho_s<0.1$), representative for the fast-ion driven TAEs. 

However, unlike previous studies performed at JET with much lower fast ion energies \cite{citrin2014electromagnetic} or in experimental results from NSTX \cite{gorelenkov2010anomalous}, the much larger amplitude of $A_{||}$ induced by the fully destabilized FI-driven modes does not lead to strong electromagnetic fluxes. Whereas the electromagnetic component of the electron transport remains low ($\chi_{EM}$(el.) $=3.1$ $\mathrm{m^2/s}$), the thermal ion one is almost negligible ($\chi_{EM}(^3\mathrm{He})$ $=0.2$ $\mathrm{m^2/s}$).
This is because the perturbed magnetic potential $A_{||}$ also exhibits a clear zonal structure in the poloidal direction, similar to the pattern for the perturbed electrostatic potential. The onset of such structures can be seen in Figure \ref{ContPlotxy_Spectra}(c) and (d) which compare the computed spatial patterns of the vector potential $A_{||}$ of the perturbed magnetic field for two cases: (1) plasma without fast ions and (2) in the presence of fast ions with a large pressure gradient, $R/L_{p_{FD}} = 16.2$. We can further quantify the impact on turbulence by computing the radial derivative of the binormal component of the zonal magnetic field perturbations, defined as $\bar{s}_{fluc} = qR_0 \partial_x B_y(k_x,0)$. Indeed, $\bar{s}_{fluc}$ is almost doubled when the TAEs are excited by fast ions ($\bar{s}_{fluc}=5.6 \times 10^{-2}$ $c_s/a$ vs. $\bar{s}_{fluc}=3.2 \times 10^{-2}$ $c_s/a$).



Whereas the coupling with electrostatic perturbations has been discussed in previous analysis \cite{di2019electromagnetic}, our results highlight the strong spatio-temporal coupling between the fast-ion driven modes and the zonal components for both field perturbations, including the electromagnetic ones.
Such novel mechanism is crucial for ITER, as a degrading electron confinement would have a negative impact as well on the thermal ions as both are strongly coupled by collisions. As a result of this complex nonlinear interaction, an improved thermal confinement and energy insulation are achieved in plasmas with a substantial population of MeV-range fast ions.

\subsection*{A promising result for ITER and future fusion devices}
Efficient heating of bulk ions was observed in recent JET fast-ion experiments in D-$^3$He plasmas with strong core electron heating from MeV-range D ions. The detailed transport analysis of these studies reveals, for the first time at JET experimental conditions, a novel mechanism of turbulence suppression in the presence of highly energetic ions and fully destabilized Alfvén eigenmodes, observed experimentally.

This promising result indicates that similar conditions might be realized in ITER and future fusion power plants, where MeV-range alpha particles provide a strong source of core electron heating \cite{stix1972heating} and can potentially destabilize TAEs \cite{gorelenkov2014energetic}. While the latter effect is usually considered as potentially detrimental for plasma confinement, the results of our studies show that this is not necessarily so. AEs can contribute to efficient ion heating through the onset of intense zonal flows, providing an improved thermal insulation of the central plasma. Thus, AE activity is not necessarily a phenomenon to be avoided in D-T plasmas, but actually one that could be purposely tailored using external actuators \cite{garcia2019active}. The identification of this novel type of for turbulence suppression has important implications for fusion research, as it holds promise to enhance the performance of future fusion devices with strong alpha particle heating, and thus could ultimately lead to an accelerated realization of commercial fusion power plants.     

We finally note that the results obtained in this paper shed light on the concept of \textit{anomalous} ion heating by alpha particles that was introduced to explain the observed increase in $T_i$ during the past full-scale D-T experiments on JET \cite{thomas1998observation}. The novel mechanism presented in this study, which includes fully destabilized TAEs triggering both electrostatic and electromagnetic zonal perturbations, highlights that this anomalous ion heating could well be identified with the suppression of microturbulence, resulting in a very effective heating of the thermal ions in plasmas with a significant population of MeV-range fast ions.	

\bibliography{MyBiblio.bib}
\bibliographystyle{unsrtmod}
	
\section*{Methods}
	
\subsection*{GENE simulation model and input parameters}
In its flux-tube version, the \textsc{Gene} code solves the nonlinear gyrokinetic Vlasov equations \cite{brizard2007foundations} coupled to Maxwell's equations on a field-aligned set of spatial coordinates (for a schematic representation, see the top left inset of Figure \ref{Tokamak}). This system of coordinates allows the exploitation of the strong anisotropy of the turbulent fluctuations in parallel and perpendicular directions to the background magnetic field. Whereas the parallel spatial direction $z$ employs a finite-difference solution technique, the perpendicular direction is treated with spectral methods, and hence the radial $x$ and binormal $y$ coordinates in the Fourier domain are generally referred as $k_x$ and $k_y$ respectively. The two velocity dimensions employed in \textsc{Gene} are the parallel velocity $v_{||}$ and the magnetic moment $\mu$. A detailed description and derivation of the model equations employed in the flux-tube version can be found in Refs. \cite{jenko2000electron,merz2008gyrokinetic}. 
	
In the present work, both perpendicular and parallel magnetic field fluctuations are computed, and the collisions are retained. In \textsc{Gene}, Maxwellian distribution functions are employed for all the particle populations, including fast D ions. For the fast ion distribution, effective density and temperature were calculated from the TRANSP distribution. The number of grid points used in the nonlinear simulations is ($n_{k_x}=256, n_{k_y}=48, n_{z}=32, n_{v_{||}}=48, n_{\mu}=64$) for 4 different particle populations, i.e.~electrons, thermal D and $^3$He ions, and fast D ions. The minimum wavenumber in the binormal direction considered is $k_{y,min}\rho_s=0.025$. The employed numerical discretization was chosen after extensive convergence tests, even in the nonlinear phase. It must be noted that in the simulations without fast particles, the velocity space has been relaxed to a less demanding numerical grid. Namely, the number of points in the $\mu$-direction has been set to $n_{\mu}=16$. Indeed, in order to accurately resolve the high-frequency resonances in fast-ion velocity space, the non-equidistant Gauss-Legendre discretization in this direction had to be increased up to $n_{\mu}=64$. Also, for a more comprehensive description of the numerical implementation techniques and adopted schemes employed in \textsc{Gene}, one can consult dedicated Ph.D. theses, e.g.~\cite{merz2008gyrokinetic}, which are available on the \textsc{Gene} website. The input parameters that have been employed in the numerical analyses are reported in \linebreak Table \ref{Table_input_parameters}.
	
\begin{table}[ht!]
	\centering 
	\caption{Employed plasma parameters in \textsc{Gene} simulations modelling JET pulse \#94701 at $\rho_{tor}=0.23$ and $t=9.5$ s.}
	\label{Table_input_parameters}
	\begin{tabular}{c c c c c c c c c}
		\hline \hline
		$\epsilon$ & $q$ & $\hat{s}$ & $T_i/T_e$ & $R/L_{n_e}$ & $R/L_{T_{e,i}}$ & $n_{\mathrm{D}}/n_e$ & $n_{^3\mathrm{He}}/n_e$ & $R/L_{n_\mathrm{D}}$\\
		\midrule
		0.31 & 1.1 & 0.63 & 1.0 & 4.50 & 10.30 & 0.43 & 0.27 & 3.70\\
		\hline \hline
		$R/L_{n_{^3\mathrm{He}}}$ & $n_{\mathrm{FD}}/n_e$ & $T_{\mathrm{FD}}/T_e$ & $\beta_e$ [$\%$] & $\nu^*$ & $B_0$ [T] & $T_{e}$ [keV] & $n_e$ [$\mathrm{m^{-3}}$] & $R_0$ [m]\\
		\midrule
		4.97 & 0.03 & 33.8 & 0.68 & $9.4 \times 10^{-5}$ & 3.68 & 4.41 & $5.18 \times 10^{19}$ & 3.00 \\
		\hline \hline
	\end{tabular}
	\caption*{Here, $\epsilon$ represents the inverse aspect ratio, $n$ the density, $R/L_{n,T}$ the normalized logarithmic density and temperature gradient, $\beta_e$ the electron-beta, and $\nu^*$ the normalized collision frequency. We also report the on-axis magnetic field strength $B_0$, the local (at $\rho=0.23$) electron temperature $T_{e}$ and density $n_{e}$, and the major radius $R_0$. The reported input parameters are common to all the numerical \textsc{Gene} simulation cases. The various cases, however, differ essentially in the fast ion pressure gradient $R/L_{p_{\mathrm{FD}}}$, whose values are displayed e.g.~in Figure \ref{HeD_DiffES}.}
\end{table}
\vspace{-2em}

\subsection*{Interpretive integrated modelling framework}
Both pulses \#94701 and \#94704 were analysed through interpretive simulations performed with the TRANSP modelling suite \cite{ongena2012numerical} coupled with external heating modules NUBEAM (NBI) \cite{pankin2004tokamak} and TORIC (ICRH) \cite{brambilla1999numerical}, and prepared with the OMFIT integrated modelling platform \cite{grierson2018orchestrating}. The interpretive analysis was based on the use of fitted profiles, including electron density and temperature - fits for both quantities were based on high resolution Thomson scattering measurements, while the temperature was additionally constrained by electron cyclotron emission data. Based on experimental measurements, the temperature of bulk ions was prescribed to be equal to the electron temperature. The equilibrium used was an EFIT reconstruction constrained by magnetics and pressure profiles, i.e. including kinetic profiles as well as the contribution of fast ions.

\subsection*{Description of experimental measurements}
Mirnov coils are used as a standard MHD diagnostic on almost all tokamak devices. The coils are installed within the vacuum vessel close to the plasma boundary and provide a measurement of the time derivative of the magnetic field. Magnetic spectrograms (Fourier decompositions of the Mirnov coil signal) can then be used to identify relevant oscillation frequencies associated with MHD activity. In JET a number of coil arrays with high frequency response are available, allowing activity in the Alfvén range to be observed. The radial localization of the modes was obtained using an X-mode reflectometer (see Supplementary information). The ion temperature profiles in this paper were obtained from Charge eXchange Recombination Spectroscopy (CXRS) measurements and electron temperature profiles from combined analysis of the electron cyclotron emission (ECE) and high resolution Thomson scattering (HRTS) diagnostics.  The density profiles were taken from HRTS measurements, with the density normalized to match the line averaged density measured by a Far Infrared interferometer. The time resolved neutron yield in JET is measured using three fission chambers, containing $^{235}\mathrm{U}$ and $^{238}\mathrm{U}$, located outside the vacuum vessel.
The time-of-flight spectrometer for rate (TOFOR) measures the energy distribution of fusion-born neutrons by based on neutron time-of-flight measurements.

\subsection*{Data availability}
The JET experimental data is stored in the PPF (Processed Pulse File) system which is a centralised data storage and retrieval system for data derived from raw measurements within the JET Torus, and from other sources such as simulation programs. These data are fully available for the EUROfusion consortium members and can be accessed by non-members under request to EUROfusion.
	
Numerical data that support the outcome of this study are available from the corresponding author upon request.

\section*{Acknowledgments}
We would like to thank Tobias Görler for providing an essential advice to ensure the correct numerical setup for \textsc{Gene} simulations reported in this paper. S.M. would also like to thank Elena de la Luna for priceless support in detailing the experimental diagnostics of JET, Yann Camenen, Xavier Garbet and Andreas Bierwage for fruitful discussions about the gyrokinetic analyses, Gerardo Giruzzi for valuable suggestions on the article strategy, and Aaron Ho for assisting the author in processing the experimental data. We sincerely thank our colleagues Matteo Baruzzo and Filomena Nave for the excellent preparation and execution of the experiments discussed in this paper.
The simulations were performed on IRENE Joliot-Curie HPC system, in the framework of the PRACE projects IONFAST and AFIETC, and on CINECA Marconi HPC within the project GENE4EP.
	
This work has been carried out within the framework of the EUROfusion Consortium and has received funding from the Euratom research and training programme 2014–2018 and 2019–2020 under Grant agreement No 633053. The views and opinions express herein do not necessarily reflect those of the European Commission.
	
\section*{Author contributions}
S.M. performed the gyrokinetic simulations and the subsequent analyses, with the fundamental support of J.G., D.Z. and S.B. The analysis shedding lights on the excitation of the fast-ion driven modes were performed by D.Z. The reported experiments were devised and jointly led by Ye.O.K., J.O., J.G. and M.N., with the key assistance of M.D. in analyzing the experimental outcomes. Crucial input data for the gyrokinetic analyses were provided by \v{Z}.\v{S}., G.S., Ye.O.K. and M.D., whereas J.E. and A.S. supported the analysis of the integrated modelling results. The manuscript was written by S.M., J.G., D.Z, Ye.O.K. and J.O. with feedback by all the authors.
	
\end{document}